\documentstyle[11pt,rotating,epsfig]{ichep}
\begin{document} 
\title{Breakthrough into the Sub-eV Neutrino Mass Range: Status of
the HEIDELBERG-MOSCOW Double Beta Decay Experiment with enriched $^{76}$Ge}
\author{A.Balysh$^b$,M.Beck$^a$,S.T.Belyaev$^{\ast,b}$,J.Bockholt$^a$,A.Demehin$^b$,
M.Eskef$^a$,D.Glatting$^a$, A.Gurov$^b$,G.Heusser$^a$,J.Hellmig$^a$, M.
Hirsch$^a$, C. Hoffmann$^a$,
 H.V. Klapdor-Kleingrothaus$^{\ast,a}$,
I.Kondratenko$^b$,D.Kotel'nikov$^b$,V.I.Lebedev$^b$,
 B.Maier$^a$, A.M\"uller$^c$, F.Petry,$^a$, E.Scheer$^a$, H.Strecker$^a$,
M.V\"ollinger$^a$, K.Zuber$^d$}
\affil{$^a$Max-Planck-Institut f\"ur Kernphysik, P.O.BOX 103980,
 69029 Heidelberg, Germany\\
 $^b$Russian Scientific Center-Kurchatov Institute, 123182 Moscow, Russia\\
 $^c$INFN Gran Sasso,67010 Assergi, Italy\\
 $^d$Institut f\"ur Hochenergiephysik,Schr\"oderstr.90, 69120 Heidelberg,
Germany
 \vspace{0.7cm}\\
\large presented by K. ZUBER}
%
%
\abstract{Recent results of the Heidelberg-Moscow double beta decay
experiment are presented. After 8.6 kg$\cdot$a of measuring time no signal is
seen for the neutrinoless decay mode. A half-life limit of
$T_{1/2}^{0\nu\beta\beta} > 5.1\cdot 10^{24}$~a is deduced which converts into
a neutrino mass limit of $\langle m_{\nu_e}\rangle < 0.68 $ eV (90\% CL).
The experiment thus is the first one penetrating into the sub-eV
range for the neutrino mass.
For the $2\nu$ mode a half life of $T_{1/2}^{2\nu\beta\beta} = (1.53\pm0.04
_{stat}\pm0.13_{sys})\cdot
10^{24}$ a is derived. More than 10000 $2\nu$ double beta
events are observed. This is the first high statistics observation of this
nuclear decay mode. Limits on
more exotic decay modes are also presented. Concerning dark matter the
experiment now gives the sharpest limits for the observation of WIMPs.}
\twocolumn[\maketitle]
\fnm{7}{Spokesmen of the collaboration}
\section{INTRODUCTION}
Neutrino physics has entered an era of new actuality. Several possible
indications of physics beyond the Standard Model (SM) of particle physics are
at present discussed: The lack of solar neutrinos, the atmospheric $\nu_\mu$
deficit and mixed dark matter models could all be explained by non-vanishing
$\nu$-masses. Recent extended SO(10) scenarios with an S$_4$ horizontal
symmetry which could explain these observations
would predict degenerate $\nu$-masses within 1-2 eV \cite{moh}. At present
$\beta\beta$-decay research may play a decisive role, since with second
generation $\beta\beta$ experiments like the Heidelberg-Moscow experiment using
large amounts of enriched $\beta\beta$-emitter material such a prediction can
be tested in very near future.

The following decay modes for double beta
decay of a nucleus $^Z_AX$ are usually considered:
\begin{eqnarray}
2\nu \beta \beta:     & ^Z_AX &\rightarrow\; ^{Z+2}_AX + 2 e^- + 2 \bar \nu_e
\\
0\nu \beta\beta:      & ^Z_AX &\rightarrow\; ^{Z+2}_AX + 2 e^- \\
0\nu \chi \beta\beta: & ^Z_AX &\rightarrow\; ^{Z+2}_AX + 2 e^- + \chi \\
0\nu 2\chi \beta\beta:& ^Z_AX &\rightarrow\; ^{Z+2}_AX + 2 e^- + 2\chi
\end{eqnarray}
A positive evidence of processes (2)-(4) would require massive
Majorana neutrinos, in addition a contribution from right-handed weak currents
is possible.
The quantity which can be extracted out of
process (2) is the effective Majorana mass given by
\begin{equation}
\langle m_{\nu_e}\rangle = \sum_i \mid U_{ei}^2 m_i \mid
\end{equation}
where $m_i$ are the mass eigenstates.
Though there exists the possibility for destructive interference,
in most grand unified models the effective mass seems to be
equal to the electron neutrino mass eigenstate \cite{lang}.

Process (3) would reflect breaking of a global (B-L) symmetry. The Majoron
$\chi$ would be the Goldstone-boson associated with this symmetry breaking.
Process (4) with the emission of two majorons seems
possible in the case of supersymmetric models \cite{moha}, and the same
spectral shape for the sum energy of the electrons
is expected if the majoron carries a
 leptonic charge \cite{cliff}.
The isotope under study at the Heidelberg-Moscow experiment \cite{bal}
 is $^{76}$Ge
in form of low-level semiconductor detectors. Further topics which can be
investigated with such detectors are, for example, charge
non-conservation \cite{bal2} and dark matter \cite{beck2}. \\

\medskip
\begin{center}
\section{THE EXPERIMENT}
\end{center}
The $Q$-value of the decay to $^{76}$Se is $Q$ = 2038.56 keV. The experiment
has 16.9 kg of Ge enriched to 86\% in $^{76}$Ge at hand (natural abundance of
 $^{76}$Ge is 7.8 \%). At present five enriched HP-detectors have been
built from this material with a total mass of 11 kg of Ge.
Table \ref{det} gives some characteristics of
the three detectors at present in regular operation at the Gran Sasso
Laboratory. The given background B is determined between 2000-2080 keV. The
next two detectors will be installed in fall 1994.
The whole
experiment is located underground in the Gran Sasso Laboratory in Italy
(shielding depth is about 3500 m.w.e.).
\begin{table}[hhh]
\small
\centering
\begin{tabular}{|c|c|c|c|}
\hline
Detector & Active mass & Enrichment& Background \\
 & [kg] & [\%] & [C/(kg$\cdot$keV$\cdot$a)]\\
 \hline
Enr.1 & 0.920 & 86 & 0.16\\
Enr.2 & 2.758 & 86 & 0.21\\
Enr.3 & 2.324 & 88 & 0.22\\
\hline
\end{tabular}
\caption{\label{det} A comparison of the different detectors used in the
experiment}
\end{table}
 The
detectors are built in a 10 cm shield containing ultrapure LC2-lead surrounded
by
another 20 cm box of very pure Boliden lead. The whole setup is plugged in an
air-free
box which is flushed with high purity nitrogen. All parts of the detector
cryostats are well selected and cleaned and have specific activities of less
than
10 $\mu$Bq/kg. To check the stability of the
experiment a calibration is done every week with a $^{228}$Th source.\\

\medskip
\section{RESULTS}
\begin{center}
{\it Neutrinoless double beta decay}
\end{center}
Fig.\ref{line} shows the region of interest of the $\beta\beta$-spectrum after
a measuring time of 8.6 kg$\cdot$a.
\begin{figure}
\caption{\label{line} The energy range between 2000-2080 keV of the spectrum
seen after 8.6 kg$\cdot$a of measurement. The dotted curve
shows the peak excluded with 90\% CL.}
\end{figure}
No line is seen
so far. For a signal hidden in the background a half-life
limit for this decay mode of $T_{1/2}^{0\nu} > 5.1 (8.6) \cdot 10^{24}$~a
 (90\% and 68\%
CL) can be derived. Using the matrix-elements given by \cite{ast} this can be
converted into a limit for the neutrino mass of
\begin{equation}
\langle m_{\nu_e}\rangle < 0.68 (0.52) \hbox{ eV} \quad 90\% (68\%) \hbox{ CL}
\end{equation}
This is the most stringent limit for a neutrino mass coming out of double beta
decay so far.

 A transition to the first excited
state would be dominated by right-handed currents and not by the mass-term like
the
ground state transition. For this transition we obtain a half-life
limit to the first excited state of
$T_{1/2}^{(0^+ \rightarrow2^+)}(^{76}{\rm Ge}) > 6.5 \cdot 10^{23}$ a (90\%
CL).
\begin{center}
{\it 2$\nu$ double beta decay}
\end{center}
In contrast to the neutrinoless decay the 2$\nu$-mode has been observed in
several
isotopes, and there is also some evidence for an observation in $^{76}$Ge.
This decay mode can give information on nuclear structure.
With the help of a Monte-Carlo simulation we subtracted all
background components which can be identified, localized and quantitatively
determined. The
remaining counts are well fitted by a 2$\nu\beta\beta$-spectrum (Fig.
\ref{2nuu}). Using a maximum likelihood
fit, we derive a half life of $T_{1/2}^{2\nu} = (1.53\pm0.04
_{stat}\pm0.13_{sys}) \cdot 10^{21}$a
after a measuring time of 49.6 mol$\cdot$a (see also \cite{bal94}). The
number of $\beta\beta$ events is about 13500 and this decay is the main
component of the measured spectrum in the region 500-1500 keV.
\begin{figure}
\caption{\label{2nuu} Remaining spectrum of enriched detector 2 after
subtraction of all known background components (solid histogram, measuring time
19.2 mol$\cdot$a). It is well
fitted by a 2$\nu$-spectrum (dashed line). Also shown is the measured spectrum
stripped of all identified peaks (dotted histogram).}
\end{figure}
\begin{center}
{\it Majoron-accompanied double beta decay}
\end{center}
For Majoron-accompanied 0$\nu\beta\beta$ decay we look by further subtraction
of the 2$\nu\beta\beta$ spectrum from the 'remaining spectrum' of Fig.
\ref{2nuu}. We obtain a half-life limit for this decay mode of
$T_{1/2}^{0\nu\chi} > 7.81 \cdot 10^{21}$ a (90 \% CL). This half-life limit
can
 be converted into a limit for the neutrino-majoron-coupling
 constant $\langle g_{\nu \chi} \rangle$, since
\begin{equation}
T_{\frac{1}{2}}^{-1} [{\rm a}^{-1}] = \mid M_{\rm GT} - M_{\rm F}
	\mid^2 F^{0 \nu \chi} \mid \langle g_{\nu \chi} \rangle \mid^2
\end{equation}
where
\begin{equation}
\langle g_{\nu \chi} \rangle = \sum_{ij} g_{\nu \chi} U_{ei} U_{ej}
\end{equation}
resulting in $\langle g_{\nu \chi} \rangle < 2.4 \cdot 10^{-4}$. For details
see \cite{beck}. The set of matrix elements given by \cite{ast} is used.
An investigation of the $0\nu2\chi\beta\beta$-mode
results in a half-life limit of $T^{0\nu\chi\chi}_{1/2} > 7.05 (8.54)
\cdot 10^{21}$ a (90 \% CL and 68 \%CL).
\begin{center}
{\it Dark matter}
\end{center}
Looking at dark matter interactions via coherent scattering of WIMPs off
Ge nuclei requires sensitivity at low energies.
By complementarily using an enriched $^{73}$Ge
detector, one would have a clear separation between spin-independent and
spin-dependent interaction ($^{73}$Ge is the only germanium isotope with spin).
After 165 kg$\cdot$d of measurement with a lower threshold of 10 keV we
obtained the exclusion plot shown in Fig. \ref{dm}.
\begin{figure}
\caption{\label{dm} Exclusion limits for WIMPs with Ge detectors (dashed area).
The increase of the sensitivity expected for some future [11] with enriched
$^{73}$Ge cryo detectors is also indicated
(curves 99 and 99.9\%). Also shown are theoretical expectations for
spin-dependently interacting WIMPs from different GUT models.
}
\end{figure}
\begin{figure}
\caption{\label{comp}
Present situation and perspectives of the most promising $\beta\beta$
experiments. Only for the isotopes shown $0\nu\beta\beta$ half life limits
$> 10^{21}$ a have been obtained. The thick solid lines correspond to the
present, 1993, situation, open bars and dashed lines to 'safe' and 'less safe'
expectations for 1999 .
}
\end{figure}
The background is lower by a factor of 5-7 than that of other Ge-experiments
dedicated to dark matter search,
therefore improving the limits for heavier WIMPs \cite{beck2,beck3}.\\

\medskip
\section{CONCLUSION and FUTURE}
Using about 10 kg of enriched material, the final goal of the
Heidelberg-Moscow experiment will
be a half-life limit of $T_{1/2}^{0\nu} > 10^{25}$a corresponding to neutrino
masses
down to about 0.1 eV. Thus it will be possible to check beyond the mentioned
SO(10)xS$_4$ model some other left-right symmetric
see-saw models. The sensitivity of running and planned experiments till the end
of the century is
shown in Fig. \ref{comp}. It is obvious that for making a significant step
beyond the limit of about 0.1 eV in reach of the present most sensitive
experiments, {\it very} large increases of set-ups are necessary (see e.g.
\cite{rag,moe})\\[0.3cm]
{\bf Acknowledgements:}\\[0.3cm]
The experiment is supported by the Bundesministerium f\"ur Forschung und
Technologie,
the State Committee of Atomic Energy of Russia and the INFN. We thank the INFN
Gran Sasso and especially Profs. P. Monacelli, E. Bellotti
and N. Cabibbo for their generous support.
%

%
%


\end{document}